\documentclass{article}
\usepackage{amssymb,amsmath}

\usepackage{PRIMEarxiv}
\usepackage{algorithm}
\usepackage{algpseudocode}
\usepackage[utf8]{inputenc} 
\usepackage[T1]{fontenc}    
\usepackage{hyperref}       
\usepackage{url}            
\usepackage{booktabs}       
\usepackage{amsfonts}       
\usepackage{nicefrac}       
\usepackage{microtype}      
\usepackage{lipsum}
\usepackage{fancyhdr}       
\usepackage{graphicx}       
\graphicspath{{media/}}     
\usepackage{float}

\title{Machine Learning Assist NYC Subway Navigation Safer and Faster
}

\author{
  Wencheng Bao \\
  \textit{Dept. of Applied Physics and Applied Mathematics}\\
  \textit{Columbia University in the City of New York}\\
  New York, NY\\
  \texttt{wb2395@columbia.edu} 
  \And
  Shi Feng\\
  \textit{Dept. of Applied Physics and Applied Mathematics}\\
  \textit{Columbia University in the City of New York}\\
  New York, NY \\
  \texttt{sf3164@columbia.edu} \\
}

\begin{document}
\maketitle

\begin{abstract}
Mainstream navigation software, like Google and Apple Maps, often lacks the ability to provide routes prioritizing safety. However, safety remains a paramount concern for many. Our aim is to strike a balance between safety and efficiency. To achieve this, we're devising an Integer Programming model that takes into account both the shortest path and the safest route. We will harness machine learning to derive safety coefficients, employing methodologies such as generalized linear models, linear regression, and recurrent neural networks. Our evaluation will be based on the Root Mean Square Error (RMSE) across various subway stations, helping us identify the most accurate model for safety coefficient estimation. Furthermore, we'll conduct a comprehensive review of different shortest-path algorithms, assessing them based on time complexity and real-world data to determine their appropriateness in merging both safety and time efficiency.
\end{abstract}


\section{Introduction}

The subway system of New York City has been used for more than 100 years[\cite{neitzel_gershon_zeltser_canton_akram_2009}]. However, New York subway system is also infamous for its unsafe. NYPD had reported 6,793 arrests in the transit system until Oct in 2022\cite{miller_2022}. Therefore, it is essential to choose the fastest and the safest route to save our life. Nowadays, people have tools such as Google Maps and Apple Maps to find the route between two subway stations, which are free and easy to use. For instance, Google Maps uses  Dijkstra's Algorithm and A* graph algorithm to generate the best route\cite{ada_2022}. However, these Apps are not able to provide the safest route. Thus, We want to include safety as a consideration in the navigation system to find the safest and most efficient route through the interchange between subways. 

The problem of predicting safety incidents can be transformed into predicting how many safety incidents will occur in the future. This is a problem of time series prediction. In 2020, Dr. Li pointed out that long and short-term memory (LSTM) has a good performance in accident prediction problems \cite{li_abdel-aty_yuan_2020}. In addition, this problem can also be seen as a queuing system problem. The Poisson process is used to predict the probability of occurrence for traffic accidents. \cite{liao_wu_yang_barth_2023}. Therefore, our plan is also to use these two algorithms and related similar algorithms for calculation. At the same time, except for starting with the traditional Single Source Shortest Path algorithm, an increasing number of path planning choices are using reinforcement learning strategies, such as Q learning, to handle path planning, such as mobile robots path planning\cite{du_hao_zhao_zhang_wang_yuan_2022}. Therefore, we also plan to test relevant policies in our project. 

In this project, We will only consider all subway lines from Columbia University to New York University. We will use all the stations between these two subway stations as a graph and try to find the optimal path. we will use Poisson Regression, Least Square, Ridge Regression, Lasso Regression, LSTM, and Gated recurrent unit (GRU) to predict future safety coefficients for each station and to use RMSE to test different performances for each algorithm. Also, we will find the shortest path using three methods: Dijkstra, Bellman-Ford, and Q-learning. We will compare the running time and performance of the three algorithms and analyze which one is the most appropriate to apply in what situations.   

\section{Datasets}

\subsection{Overview of NYC Subway Data}
The dataset of New York subway systerm comes from NYC Open Data Subway Stations\cite{new_york_city}, which represents the connections among subway stations. Based on the information provided, all the possible routes to travel from Columbia University (1 train 116th St Station) to New York University (R,W train 8th St Station) from NYC subway system can be gained. We will construct the graph by using Networkx with the following columns stations, train, previous stop, next stop, and time. We will assume that there is no in-station transfer walking time and technical delays.

\begin{figure}[h!]
\begin{center}
\includegraphics[scale=.5]{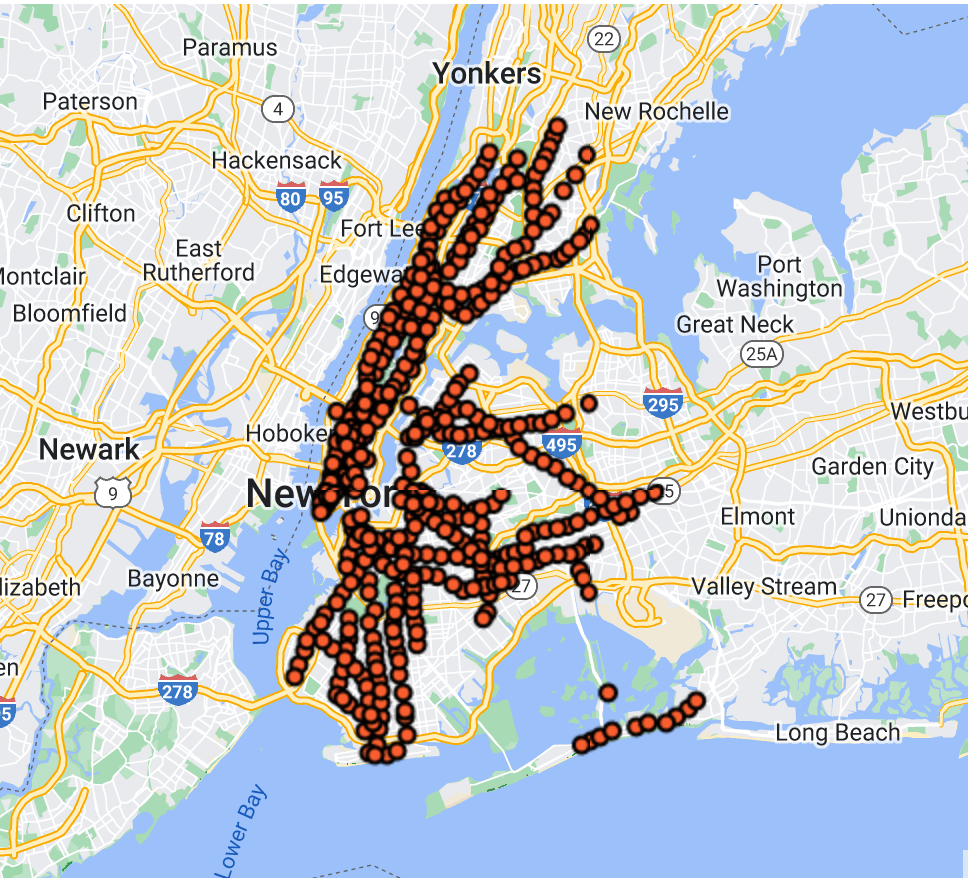}
\caption{Sample NYC Subway Route \cite{new_york_city}} 
\end{center}
\end{figure}

The goal is to focus on the traveling time, from one station to another, and the methods to connect stations in the paper. We will create a net flow chart and generate the shortest path from Columbia to New York University.

\subsection{NYC Open Data from NYPD}

We are collecting data from NYC Open Data from NYPD\cite{(nypd)_2023}. This data set contains the times and locations of crimes reported in NYC which from 2006 to the end of 2019. We will focus on the time and coordinates of each incident in the dataset. Based on the latitude and longitude of all NYC subway stations and by searching for the number of incidents that occurred within 8km of that coordinate for each station from 2018 to 2019, we will obtain a univariate coordinate plot with time as the x-axis and the number of accidents as the y-axis. The goal is to predict whether the number of accidents increases or decreases as time increases.

Considering that the frequency of accidents near each station is different, we will use the last 5 time points when accidents occur as the test set and the rest of the time points as the training set.

\section{Methodology of Safety Prediction}
\label{sec:others}

\subsection{Poisson Regression}
Poisson regression belongs to a form of generalized linear model (GLM). In the problem of safety prediction, we consider that the data satisfies smoothness (x-axis time interval of 1), independence (each event occurrence is independent of each other), and generality (small frequency of occurrence, i.e., low probability). Therefore, we will do this by assuming that the frequency $y$ of accident occurrences fits the Poisson distribution. Also, it is assumed that the logarithm of the expectation can be modeled by a linear combination of unknown parameters.

In the previous data set part, we have already denoted the $x$ represents the time and $y$ represents the number of accident. Therefore, the model form of Poisson regression is $log(E[y | x]) = \alpha + \beta' x = \theta' x$ where $\alpha \in R, \beta \in R^n$, $ \theta$ construct with $\alpha$ and $\beta$. Thus, $E[y | x] = e^{\theta' x}$. By bringing $E[N(t)] = E[y | x] = e^{\theta' x}$ to Probability density function of Poisson distribution. We got:$P_y(t) = P(y|x, \theta) = \frac{(e^{\theta' x})^y}{y!}e^{- \theta' x} = \frac{e^{\theta' xy}}{y!}e^{- \theta' x}$. Then, applying maximum likelihood estimation (MLE), we can find the Joint probability: $p\left(y_1, \ldots, y_m \mid x_1, \ldots, x_m ; \theta\right)= L(\theta | x, y)  = \prod_{i=1}^m \frac{e^{y_i \theta^{\prime} x_i} e^{-e^{e^{\theta'} x_i}}}{y_{i} !}$. Then By applying the logrithm equation to the above joint distribution:

\begin{equation}
log(L(\theta | x, y)) = \sum_{i = 1}^m (y_{i}\theta x_i - e^{\theta' x_i} - log(y_i !)) = \sum_{i=1}^m (y_i \theta' x_i - e^{\theta' x_i})
\end{equation}

By finding the extreme value of equation (1), we can find the optimal value for $\theta$ to accomplish the Poisson Regression.

\subsection{Linear Regression}
In the data set, there exists $\{(x_1, y1),(x_2,y_2),...,(x_n, y_n)\}$. It is going to use $f(x) = \beta x$ to fit those data points. And by using the form of using ordinary least square (OLS), we got $\beta = (X^TX)^{-1}X^TY$. Considering the covariance matrix of $\beta$ is $\sigma^2 (X^TX)^{-1}$, which means the variance of $\beta$ may large. This may cause the regression coefficients obtained are meaningless. In order to solve the problem, we will also use both Ridge and Lasso regressions, which are add additional L2 and L1 parametric regularization. For solving the version for lasso, we have to apply soft thresholding. Then We are going to write the equation with KKT multiplier $\lambda$:

\begin{equation}
\text{For Ridge}: J(\beta) = \sum(Y - X\beta)^2 + \lambda ||\beta||^2_2 = Y^TY - Y^TX\beta - \beta^TX^TY + \beta^TX^TX\beta + \lambda\beta^T\beta
\end{equation}

Therefore, we got $\beta_{Ridge} = (X^TX + \lambda I)^{-1}X^TY$.
\begin{equation}
\text{For Lasso}: J(\beta) = \sum(Y - X\beta)^2 + \lambda ||\beta||_1 = Y'Y + \sum_{j=1}^n(\beta_j^2 -2Y'X_j\beta_j + \lambda|\beta_j|)
\end{equation}
Therefore, we got $\beta_{Lasso} = Y'X_j$ when $\beta_j =0$, $Y'X_j + \lambda/2$, when $\beta_j< 0$ and $Y'X_j - \lambda/2$ when $\beta_j > 0$
\subsection{Recurrent Neural Network}
In the Recurrent Neural Network, we are going to use Long Short Term Memory (LSTM) and Gated Recurrent Unit (GRU). Since the dataset grows according to time, we believe that the use of time series correlation algorithms will work well for the problem of security prediction. The general structure for LSTM and GRU shows below:
\begin{figure}[h!]
\begin{center}
\includegraphics[scale=.4]{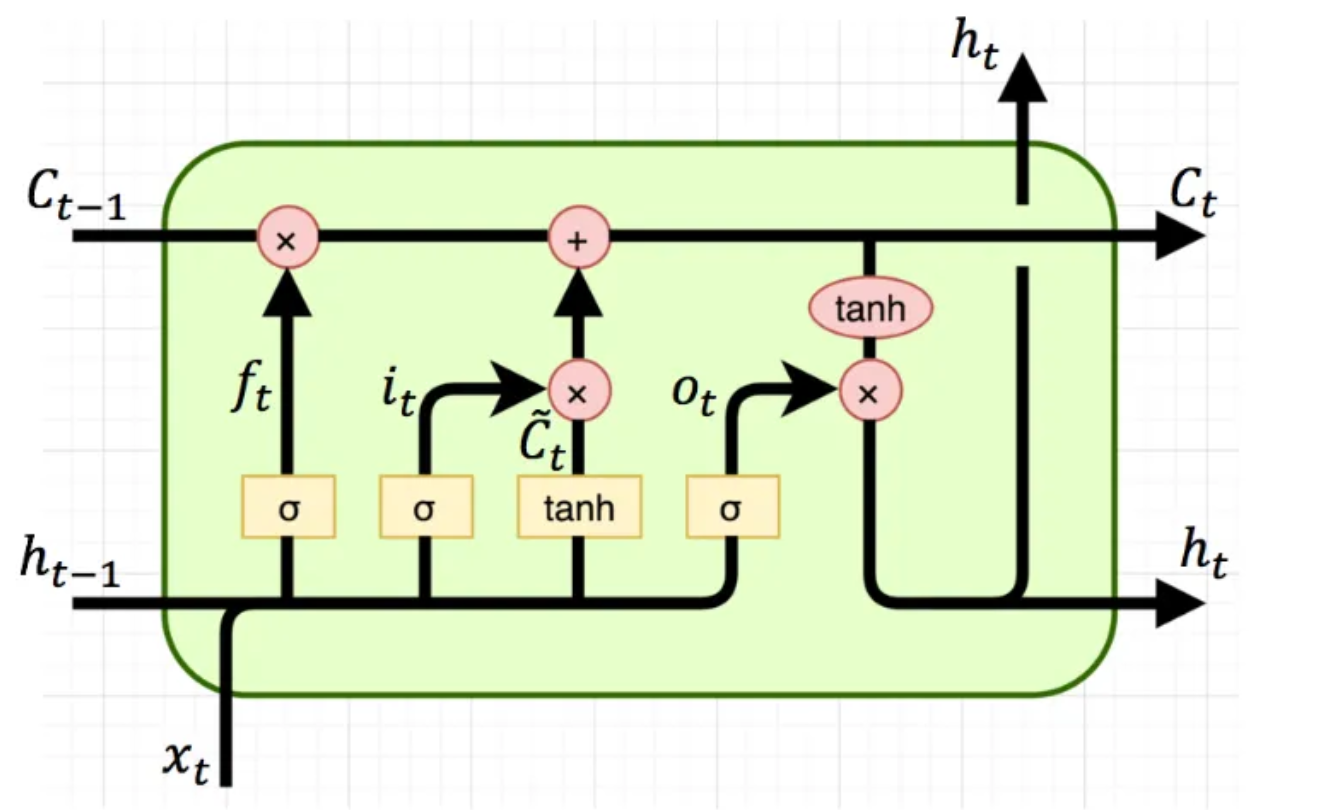}
\includegraphics[scale=.8]{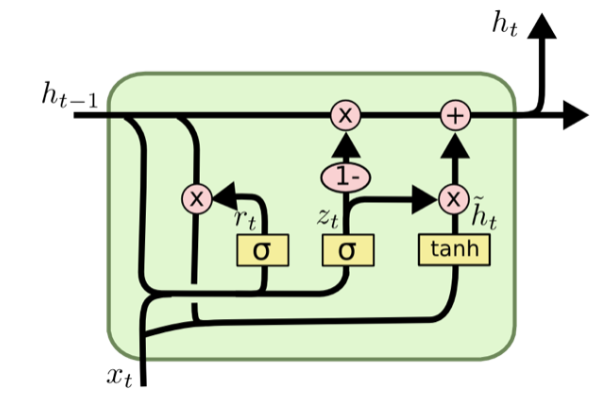}
\caption{LSTM \& GRU \cite{mani_2019}} 
\end{center}
\end{figure}

Two main channels exist in the LSTM to maintain memory. There exists a short-term memory $h$, and $h$ will be maintained for nonlinear operation. Long-term memory $C$, $C$ will maintain linear operation. We want to train to obtain the long-term memory of useful information. Meanwhile, the core of the LSTM is its state.\\
\\
The first thing for LSTM is to decide which information needs to be dropped or kept. This step needs to be handled by forget gate $(f_t = \sigma (W_f \dot [h_{t-1}, x_t] +b_f))$ by sigmoid unit. The sigmoid unit will check $h_{t-1}$ and $x_t$ to output 0 or 1, where 0 means discard, 1 means reserve. After that, need to use output gate to decide which information needs to be updated in the state. Then, we need to use tanh to obtain a new state $\hat C_t$ with $h_{t-1}$ and $x_t$: $i_t = \sigma (W_i \dot [h_{t-1}, x_t] +b_i))$ and $\hat C_t = tanh (W_c \dot [h_{t-1}, x_t] +b_c))$. Therefore, we got the new state $C_t = f_t * c_{t-1} + i_t * \hat C_t$. And finally, by using the output gate, the output for this RNN will follow:
\begin{equation}
new_t = \sigma (W_{new} \dot [h_{t-1}, x_t] +b_{new})) \quad C_t = new_t * tanh(C_t)
\end{equation}
\\
The item in GRU is similar to the LSTM. The difference for the GRU is it only contains Reset Gate and Update Gate, and do not contains other gate like for. In the above figure, which is $r_t$ and $z_t$. The $r_t = \sigma (W_r \dot [h_{t-1}, x_t])$ represents the reset gate, and $z_t =  \sigma (W_z \dot [h_{t-1}, x_t])$ represents the update gate. As same as what we did in LSTM, we will get:
\begin{equation}
\hat h_t = tanh(W \dot [r_t * h_{t-1}, x_t]) \quad h_t = (1-z) \odot h_{t-1} + z \odot \hat h_t
\end{equation}

\section{Methodology of Route Navigation}

In order to navigate the shortest path, we define $x1, x2, x3,...$ as the name of the station, the $s1, s2, s3,...$ is defined as safety coefficient, and $t1, t2, t3,..$ shows time coefficient between two stations. We are going to construct the Multi-objective Integer Programming into a normal Integer Programming problem:

$$
min \quad z = \sum_i  x_{ij} s_{ij} t_{ij}\\
$$
$$
\text{Subject to:} \quad \sum_{j=1} x_{ij} - \sum_{k=1}x_{ki} = \begin{cases}1&\text{if} \  i=s\  \text{(source)} \\ 0&\text{otherwise}\\ -1&\text{if} \  i=s\  \text{(sink)}\end{cases}, \quad \text{for} \quad  x_{ij} >= 0
$$
$x_{ij}$ means the edge from station $x_i$ to station $x_j$, and $x_{kj}$ means the edge from station $x_k$ going back to station $x_i$. If $x_i$ is a source, then $x_{ki}=0$ so $\sum_{j=1} x_{ij} - \sum_{k=1}x_{ki} = 1$ and vice versa for $x_i$ is a sink. For all $j$ and $k$, if $x_i$ is neither a source nor a sink, we set $ x_{ij}= 1$. Thus, we can construct the graph with node, edges, and weight of edges and nodes.

\subsection{Single-Source Shortest Path (SSSP)}
Following with Google or Apple map, we will use the Dijkstra to solve the above optimization equation. We also apply Bellman-Ford's to compare. Considering two algorithms have space complexity of $O(|V|)+O(|V|)=O(|V|)$ space complexity. We will only focus on the time complexity and running time.

For Dijkstra, $|V|$ stands for the number of nodes so adding all nodes to $Q$ takes $O(|V|)$. Then, to remove the nodes with the smallest distance also takes $O(|V|)$. For each iteration of looping, it takes another $O(|V|)$and one node is deleted in each loop. Thus, total time complexity becomes $O(|V|)+O(|V|)*O(|V|)=O(|V|^2)$.
Also,the Fibonacci Heap can improve the time complexity of Dijkstra's algorithm to $O(|E|+|V|log|V|)$, where $|E|$ stands for the number of edges.
\begin{algorithm}
\caption{Dijkstra's Algorithm for Shortest Path}\label{alg:cap}
\textbf{Input:} Graph \( G = (V,E) \) and distance \( s*t \) of each edge \\
\textbf{Output:} A list of nodes
\begin{algorithmic}
\State Set \( \text{dist}[116\text{th}] = 0 \), \( \text{dist}[v] = \infty \) and \( \text{previous}[v] = \text{undefined} \) for all vertices \( v \)
\State Create a list \( Q \) with all nodes
\While{Q is not empty}
    \State Find the node \( u \) with the smallest distance, and remove \( u \) from \( Q \)
    \For{each neighbor \( v \) of \( u \)}
        \If{\( \text{dist}[u] + \text{dist}[u, v] < \text{dist}[v] \)}
            \State \( \text{dist}[v] = \text{dist}[u] + \text{dist}[u, v] \)
            \State Set \( \text{previous}[v] = u \)
        \EndIf
    \EndFor
\EndWhile
\State \textbf{return} previous
\end{algorithmic}
\end{algorithm}

For Bellman-Ford's, we need to relax all the $|E|$ edges for $|V|-1$ time in total. Therefore its time complexity becomes $O((|V|-1)*|E|)=O(|V||E|)$ time. The pseudo-code shows below:

\begin{algorithm}
\caption{Bellman-Ford's Algorithm for Shortest Path}\label{alg:cap}
\textbf{Input:} Graph \( G = (V,E) \) and distance \( s \times t \) of each edge \\
\textbf{Output:} A list of nodes
\begin{algorithmic}
\State Set \( \text{dist}[116\text{th}] = 0 \), \( \text{dist}[v] = \infty \) and \( \text{previous}[v] = \text{undefined} \) for all vertices \( v \)
\For{\( i = 1 \) to \( N-1 \)}
    \For{each edge \( (u,v) \) in \( E \)}
        \If{\( \text{dist}[u] + \text{dist}[u, v] < \text{dist}[v] \)}
            \State \( \text{dist}[v] = \text{dist}[u] + \text{dist}[u, v] \)
            \State Set \( \text{previous}[v] = u \)
        \EndIf
    \EndFor
\EndFor
\For{each edge \( (u,v) \) in \( E \)}
    \If{\( \text{dist}[u] + \text{dist}[u, v] < \text{dist}[v] \)}
        \State \textbf{return} "Negative cycle exists"
    \EndIf
\EndFor
\State \textbf{return} previous
\end{algorithmic}
\end{algorithm}

\subsection{Reinforcement Learning Shortest Path}

We mainly use the Q-learning method to search for the shortest path. This algorithm is different from previous optimization algorithms. It is an algorithm for model free. It is an action-state based algorithm that iteratively updates the q value of each action state to determine the optimal path between two vertexes. Because the core is to update Q-tables, our table is composed of reward tables. If there is connectivity between node 1 and node 2, then $[\text{Node 1}] [\text{Node 2}] =-1 * \text{safety factor * time}$. If there is no connectivity between Node 1 and Node 2, then $[\text{Node 1}] [\text{Node 2}]=-99$. Any $[\text{node}] [\text{ultimate goal}]=100$. We will update this Q Table until we reach the endpoint.

\begin{algorithm}
\caption{Reinforcement Learning: Q-Learning}\label{alg:cap}
\begin{algorithmic}
\State Initialize \( Q(s,a) \) (similar to a reward table).
\For{each episode}
    \State Choose an action \( a \) from state \( s \) using a policy derived from \( Q \), e.g., the Epsilon-Greedy algorithm.
    \State Take action \( a \), and observe the resulting reward and the next state \( s' \).
    \State Update \( Q \) as:
    \State \( Q(s,a) \leftarrow Q(s,a) + \text{learning rate} \times [\text{reward} + \text{discount rate} \times \max_{a'}Q(s',a') - Q(s,a)] \)
    \State Set \( s \leftarrow s' \).
\EndFor
\State Continue until the NYU station is reached.
\end{algorithmic}
\end{algorithm}

\section{Result}

The following graph shows average RMSE safty score for each station from 116th to 23rd:

\begin{figure}[h!]
\begin{center}
\includegraphics[scale=.68]{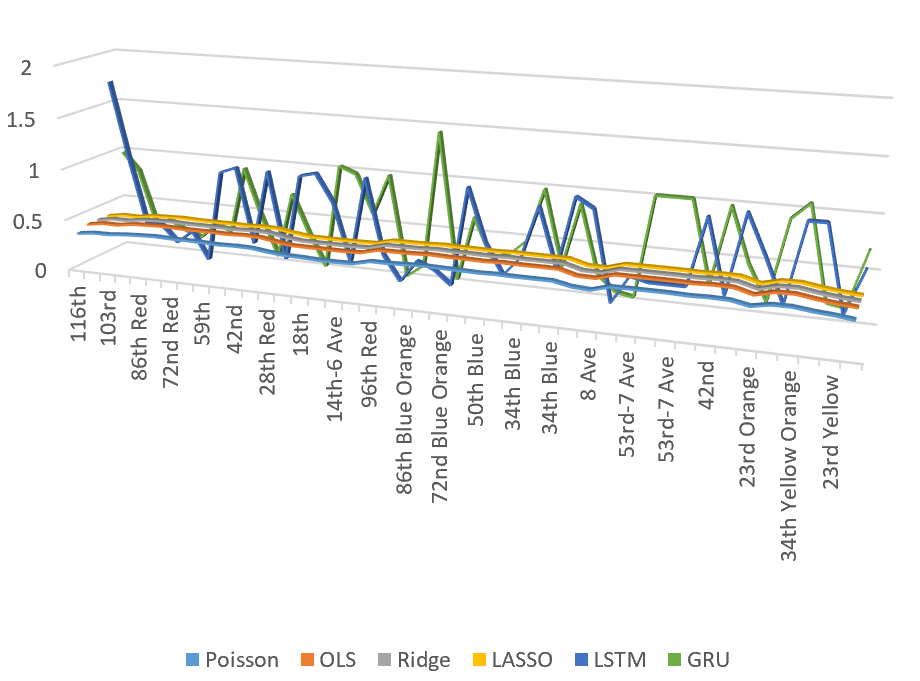}
\includegraphics[scale=.65]{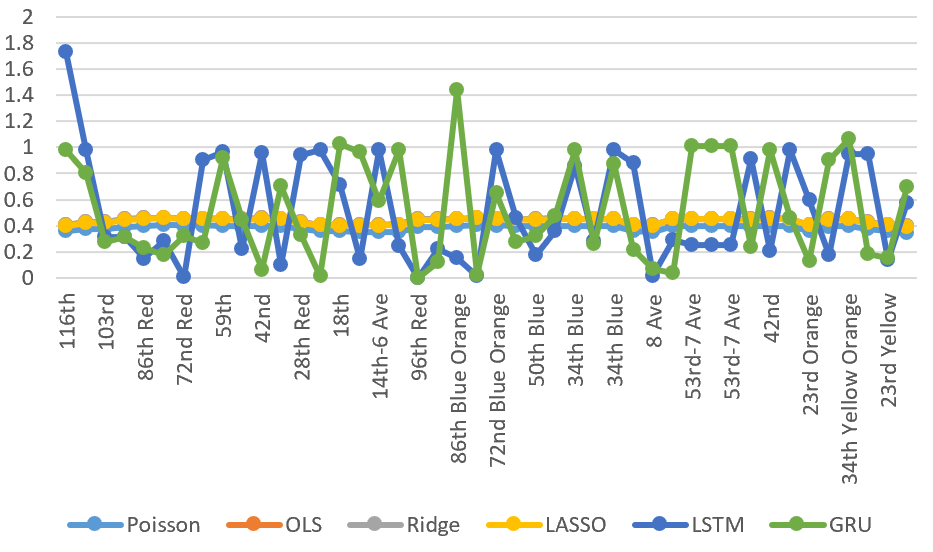}
\caption{RMSE for All Station} 
\end{center}
\end{figure}

The average RMSE scores for  Possion Regression, Least Square, Ridge Regression, Lasso Regression, LSTM, and GRU are: 0.385, 0.439, 0.439, 0.436, 0.519, and 0.522, which shows the Poisson regression has the lowest RMSE value, which means it contains the best performance. But we cannot assume the Possion is the best model. Because from the Figure 2, we can find the Poisson regression doesn't show the big difference between each stations.Hence, using Poisson Regression will make it difficult to distinguish which station is more dangerous and which station is safer. This might because in the test set, most of the frequency of accidents is concentrated between 0 and 1. Thus, this may be the reason why Poisson regression can get better RMSE results. However, in real life, we would recommend to use LSTM model since it has significant learnability for different station and also has a lower RMSE score compared to GRU.

For find shortest path, running time for Dijkstra, Bell-man-Ford and Q learning are 0.00467s, 0.00467s, and 59s. And all of the algorithms return the same fastest and safest path as: 
'116th', '110th', '103rd', '96th Red', '86th Red', '79th', '72nd Red', '66th', '59th', '50th Red', '42nd', '34th Yellow Orange', '28th Yellow', '23rd Yellow', 'Union Sq', '8th'.

Therefore, the Bellman-Ford will have the fastest running time for this situation. But by considering the time complexity, the Bellman-Ford is $O(|V||E|)$ and the Dijkstra is $O(|E| + |V|log|V|)$, we might analysis that there are more edges than vertexes, which same as fact. In reality, we will recommend Dijkstra more because of its lower time complexity.

\section{Disscussion}
In the issue of safty prediction, we recommend using LSTM. Because the dataset is a univariate problem, we believe that there should not be a linear causal relationship between time and accidents. Therefore, we prefer models that can reflect change and learning more. LSTM based on RNN also conforms to the current state of art.

Although we previously suggested using Dijkstra instead of Bellman Ford due to time complexity considerations. But if there is negative weight on the edge, we can only use Bellman Ford. Meanwhile, only Bellman Ford can determine whether there is a negative weight loop. Also, although Q-learning takes 59s to compute. But this does not mean that it is a bad algorithm. If we know the distance of nodes, edges and weights of the map, the Dijkstra or Bellman-Ford will find the shortest path efficiently. But what if there are wrong points on the map? What if the safe prediction is 0 or even negative. In reality there are many black box problems, reinforcement learning although it is possible not to find the optimal solution, but should be able to find a relatively good solution. We can then use reinforcement learning with model free policy. By using the strategy to continuously search for information in the graph, update the value function, and eventually can find the shortest path.

In the next step, we will more focus on how to use reinforcement learning to find the shortest path. We hope to find more valuable rewards to help us update policy. In addition, we would like to try some other reinforcement learning models, such as Off policy Actor critical, which approach the policy by establishing Policy Network and Value Network. Based on our judgment on the safety issue, we will use LSTM cells to build this two layer feed-forward neural network.

\bibliographystyle{unsrt}  
\bibliography{references}  

\end{document}